\documentclass[conference]{IEEEtran}
\IEEEoverridecommandlockouts
\usepackage{cite}
\usepackage{amsmath,amssymb,amsfonts}
\usepackage{algorithmic}
\usepackage{graphicx}
\usepackage{textcomp}
\usepackage{xcolor}
\usepackage{todonotes}
\usepackage[hidelinks, 			
breaklinks=true		
]{hyperref}	
\def\BibTeX{{\rm B\kern-.05em{\sc i\kern-.025em b}\kern-.08em
		T\kern-.1667em\lower.7ex\hbox{E}\kern-.125emX}}
\begin{document}
	
	\title{A Light-weight and Unsupervised Method for Near Real-time Behavioral Analysis using Operational Data Measurement
		
	}
	
	\author{\IEEEauthorblockN{Tom Richard Vargis}
		\IEEEauthorblockA{\textit{ScaDS.AI Dresden/Leipzig} \\
			\textit{Technische Universität Dresden}\\
			Dresden, Germany \\
			tom\_richard.vargis@tu-dresden.de}
		\and
		\IEEEauthorblockN{Siavash Ghiasvand}
		\IEEEauthorblockA{\textit{ScaDS.AI Dresden/Leipzig} \\
			\textit{Technische Universität Dresden}\\
			Dresden, Germany \\
			siavash.ghiasvand@tu-dresden.de}
	}
	
	\maketitle
	
	


\section*{}
\label{sec-motivation}
Monitoring the status of large computing systems is essential in order to identify unexpected behavior and improve their performance and up-time.
However, due to the large scale and distributed design of such computing systems as well as the large number of monitoring parameters, automated monitoring methods should be applied.
Such automatic monitoring methods should also have the ability to adapt themselves to the continuous changes of the computing system.
In addition, they should be able to identify behavioral anomalies in useful time, in order to perform appropriate reactions.

Majority of the existing automated monitoring tools are rule-based or highly dependent on manual adjustments~\cite{ref_ispdc2019}.
Power awareness of HPC system jobs helps prevent temperature hot-spots in the system and further failures~\cite{ref_Classifyingjobs}.
The machine learning based tools are either purposely built for certain computing systems which cannot be applied  directly to other systems, or are highly resource intensive thus, do not scale with the growing size of computing systems~\cite{ref_Experiencereport2022}.
Furthermore, the lack of labeled and recent operational measurement datasets makes the creation and comparison of supervised machine learning based approaches significantly challenging~\cite{ref_systemfailure}.
To the best of our knowledge, this is the first work which proposes a general light-weight and unsupervised method for near real-time anomaly detection using operational data measurement on large computing systems.

The proposed method has been successfully applied on operational data of Taurus\footnote{https://doc.zih.tu-dresden.de/jobs\_and\_resources/hardware\_overview/} HPC cluster which is collected using MetricQ~\cite{ref_metricq2019}.
A proof-of-concept realization of the proposed method is implemented using Python and Keras API\footnote{https://keras.io/}.
To facilitate the reproducibility of this work, the source code and sample data are publicly available~\cite{ref_gitcode}.

\section*{}
\label{data-and-model}
Changes in system’s behavior are reflected in its operational data measurements (parameters).
Processing all existing operational parameters of large-scale and high-performance computing systems (HPCs) in order to identify anomalies is practically impossible due to high overhead and long response time.
Correlation between these operational parameters is the key to address this challenge~\cite{ref_lessons2016}.
This work proposes an unsupervised approach to utilize the energy consumption and temperature measurements of the computing nodes in order to detect abnormal system behavior.

On Taurus each computing node has two CPUs and due to the design of the cooling pipeline, there are different correlations between CPUs of coupled neighboring nodes.
In total five parameters of each computing node is collected.
Power consumption of each CPU, temperature of each CPU, and the total power consumption of the computing node.
It is important to note that in contrast to energy consumption measurements which instantly reflect operational changes of each computing node, the temperature sensors reflect such changes with a short delay.

The streamed monitoring data is down-sampled to 10-second buckets.
Normalization of the data using MinMax or Standard scaler proved to play a significant role in the accuracy of predictions.
Autoencoder model was chosen for the unsupervised learning process which adapts itself to the dynamic behavior of the computing system.

In order to achieve near to real-time response time with low performance overhead, a simplified neural network using the Long Short-Term Memory (LSTM) neural network cells were designed in the Autoencoder model. 
The simplest model was defined initially with 1 encoder and decoder layer each with a bridge in between. However, this did not yield much accurate results. It improved with increasing number layers and a model  with 3 LSTM encoder layers followed by an encoder-decoder bridge and 3 LSTM decoder layers was able to achieve high accuracy without much loss of performance. Adding more layers were only seen to make the model more complex.
The proposed model requires as low as 50 epochs for each training process. Every four hours, the training process is repeated and the predictions are made continuously.
Progressive learning enables the model to adapt itself to system's behavior and optimally fit the data.

The maximum error value in predictions from the current 4-hour interval is defined as the threshold for the next interval.
The threshold value is calculated for each feature separately and is updated every four hours.


The data is fed into the model in groups of four rows (40 seconds) which was achieved using a moving window.
Consequently, the predictions were made in groups of four.
Therefore, a prediction from a particular timestamp have four different values in each group.
The average of all predictions corresponding to a particular timestamp is calculated as the final prediction.
Although this process could be performed using a single row , the multi-row approach offered significantly higher accuracy. The choice of 4 rows here is arbitary. Based on testing the average of four rows of prediction provides good accuracy. And using larger window sizes might lead to further delay and performance degradation.

The length of the moving window as well as the training intervals in this work are empirical choices.
These hyperparameters should be adjusted according to each particular use case.
In order to further optimize the proposed model, the hyperparameters domain was further explored using KerasTuner\footnote{https://keras.io/keras\_tuner/} to find the best possible values.
According to KerasTuner the best model should have one layer less than the proposed model.
However, since the accuracy slightly dropped for the new model, the original model was maintained.

The correct selection of features according to their correlation is one of the deciding factors in this work.
Using the correct selection of features in this work, the number of trainable parameters in the proposed model were reduced to less than 68,000.
Which in turn, significantly improved the performance of the model and enables a close to real-time identification of anomalies.
Experiences done on Taurus operational data shows that the proposed model quickly learns the behavioral patterns and optimally fits to the data.

With minimal training, the model provides quick predictions with an approximate accuracy of {96\%}.
However, currently any prediction with an error value above the threshold is categorized as an anomaly.
Therefore, the decision mechanism should be improved such that a cross-features decision would be made which in turn will also reduce the number of false positives.
In addition, the definition and the update-interval of feature thresholds should be fine-tuned to improve the final results.
In contrast to syslog analysis~\cite{ref_ispdc2019}, the selection of nodes based on vicinity perspective did not have a significant impact on the predicted anomalies.
Further analysis of this behavior is also planned as part of the future works.

\section*{}
\subsubsection*{Acknowledgements}
\label{acknowledgements}
This work was supported by the German Federal Ministry of Education and Research (BMBF, 01/S18026A-F) by funding the competence center for Big Data and AI "ScaDS.AI Dresden/Leipzig".
The authors gratefully acknowledge the GWK support for funding this project by providing computing time through the Center for Information Services and HPC (ZIH) at TU Dresden on HRSK-II.

%
%
%
%

\end{document}